\documentclass[12pt]{article}
\usepackage{amsfonts} 
\usepackage{graphicx}
\usepackage{xcolor}
\usepackage{siunitx}
\usepackage[export]{adjustbox}
\usepackage[labelformat=simple]{subcaption} % parenthesis on subfigure reference
\usepackage[square, numbers, sort&compress]{natbib}

\usepackage[a4paper,            left=0.7in,right=0.7in,top=1in,bottom=1in,%
            footskip=.25in]{geometry}

\usepackage{authblk}
\title{\textbf{Relaxing constraints on data acquisition and position detection for trap stiffness calibration in optical tweezers}}
\author[1]{Bruno Melo}
\author[1]{Felipe Almeida}
\author[2]{Guilherme Tempor\~{a}o}
\author[1]{Thiago Guerreiro}
\affil[1]{Department of Physics, Pontifical Catholic University of Rio de Janeiro, Rio de Janeiro 22451-900, Brazil}
\affil[2]{Center for Telecommunications Studies, Pontifical Catholic University of Rio de Janeiro, Rio de Janeiro 22451-900, Brazil}    
\date{April, 2020}

\begin{document}

\maketitle

\newcommand\myeqa{\stackrel{\mathclap{\normalfont\mbox{(i)}}}{\approx}}
\newcommand\myeqb{\stackrel{\mathclap{\normalfont\mbox{(ii)}}}{=}}
\newcommand\myeqc{\stackrel{\mathclap{\normalfont\mbox{(iii)}}}{\approx}}
\newcommand\myeqd{\stackrel{\mathclap{\normalfont\mbox{(ii)}}}{\approx}}
\newcommand\myeqe{\stackrel{\mathclap{\normalfont\mbox{(v)}}}{\approx}}
\newcommand\myeqf{\stackrel{\mathclap{\normalfont\mbox{(vi)}}}{\approx}}
\newcommand\myeqg{\stackrel{\mathclap{\normalfont\mbox{(i)}}}{=}}

\begin{abstract}
Optical tweezers find applications in various fields, ranging from biology to physics. One of the fundamental steps necessary to perform quantitative measurements using trapped particles is the calibration of the tweezer's spring constant. This can be done through power spectral density analysis, from forward scattering detection of the particle's position. In this work we propose and experimentally test simplifications to such measurement procedure, aimed at reducing post-processing of recorded data and dealing with acquisition devices that have frequency-dependent electronic noise. In the same line of simplifying the tweezer setup we also present a knife-edge detection scheme that can substitute standard position sensitive detectors.
\end{abstract}

\section{Introduction}
Optical tweezers were conceived as tools capable of harvesting radiation pressure to hold and manipulate tiny objects \cite{Ashkin1970, Ashkin1986}. Throughout the past three decades, optical trapping has encountered a vast number of applications in different research fields \cite{Polimeno2018}. They might, for instance, be used in biology to measure physical properties of cell membranes \cite{Nussenzveig2017, Dao2003} or manipulate living microorganisms \cite{Zhang2019, Zhong2013}; in chemistry, to build and trap single molecules \cite{Liu2018, Hu2019}, and in physics, as precise force sensors \cite{Monteiro2017, Hebestreit2018} or to achieve quantum ground state cooling  \cite{Delic2020, Tebbenjohanns2020}. To perform all of these tasks, it is necessary to quantitatively know the force exerted by the light beam on the trapped particle. 

While the behaviour of the force as a function of particle displacement can be predicted for different trapping beams \cite{Liu2012, Jiang2011, Zhan2003}, its numerical value is sensitive to a number of experimental variables, such as the medium's viscosity \cite{Barry2018} and spherical aberrations \cite{XinCheng2001, Theofanidou2004}. This often makes it hard to exactly calculate the relevant forces from first principles \cite{Dutra2012, Dutra2014}. Instead, these are experimentally measured during the tweezer calibration. For a Gaussian beam the trapping potential is harmonic and one of the goals of calibration is to find the trap's spring constant.

Due to randomness introduced by the medium in which the trapped particle finds itself, the spring constant is usually obtained by measuring statistical quantities associated to the particle's position. Among these quantities there are the autocorrelation function \cite{Alves2012} and the power spectral density \cite{BergSorensen2004} of the particle's position. Whichever quantity is chosen, one needs to be able to measure the particle's displacement as a function of time. This can be done in a number of ways. 

For example, video imaging \cite{Gibson2008} is useful when dealing with multiple trapped particles in holographic tweezers \cite{Drobczynski2012}. 
The detection of the forward and back-scattered light on the other hand enables measurements of the position at higher sampling frequencies \cite{Huisstede2005, Pralle1999}. 
In this paper, we will focus on the use of forward-scattering detection to measure the power spectral density of a trapped particle's position. 
In this type of setup it is standard to collect the light scattered by the bead using an objective lens and direct it onto a position sensitive detector. The detector outputs three signals as a function of time: $X$ and $Y$ - proportional to the detected light power and the position in which the light hits the detector - and $S$, proportional solely to the detected power. For small displacements, the beam deviation is proportional to the radial particle position, which can therefore be obtained by dividing $X$ and $Y$ by $S$, aside from a constant. Once this operation is performed for each signal sample, it can be used to calculate the power spectral density and consequently find the desired spring constants after fitting the data to a Lorentzian function and using the proportionality relation between the trap stiffness and the Lorentzian's corner frequency \cite{Jones2015}. %Due to this linear relation, the terms ``corner frequency" and ``trap stiffness" will be used interchangeably.

We propose simplifications to this procedure which reduce the standard hardware and computational requirements. First, we show that under reasonable approximations, the radial particle position can be directly obtained from the signals $X(t)$ and $Y(t)$, making the step of dividing signals at each sampling unnecessary. Next we show that excessive electronic noise in the detection channels can be accounted for by adapting the function used to fit the measured power spectral density. This generalizes a method proposed in the context of video imaging \cite{Horst2010}, and allows for the use of inexpensive data acquisition devices. Finally, still motivated by the interest in reducing the cost and complexity of an optical tweezer setup \cite{Smith1999, Candia2013}, we propose a knife-edge method that aims to substitute a position sensitive detector by regular power-sensing silicon detectors. These detectors also present the advantage of having an increased bandwidth which can be explored in statistical mechanics experiments involving trapped particles \cite{Grimm2012}. %All of our claims are experimentally verified and used to measure the spring constant in the radial direction of a custom-built optical trap. 

In the next section, we introduce the theory behind the proposed calibration procedure and the above-mentioned knife-edge detection system. We then describe our experimental apparatus and use it to verify both the calibration as well as the knife-edge method. We do that by measuring the power spectral density of the particle's motion and comparing it to the standard calibration of an optical tweezer. We close with a brief discussion and the conclusions of this work.

\section{Theory: data analysis and knife-edge detector}
In what follows, $\langle A(t)\rangle$ is the time average of $A(t)$ and $\delta_A(t)$ is the instantaneous deviation of $A(t)$ with respect to $\langle A(t)\rangle$. We start by taking a closer look at the division of $X(t)$ by $S(t)$, which yields the particle position in the $x$ direction when using standard forward scattering detection \cite{Jones2015},
\begin{eqnarray}
\label{eq:approximation1} 
    \frac{X(t)}{S(t)}&=&\frac{\langle X(t)\rangle+\delta_X(t)}{\langle S(t) \rangle + \delta_S(t)}\approx\frac{\langle X(t)\rangle+\delta_X(t)}{\langle S(t) \rangle}\left(1-\frac{\delta_S(t)}{\langle S(t)\rangle} \right)=\nonumber\\
    &\approx&\frac{\delta_X(t)}{\langle S(t) \rangle}\left(1-\frac{\delta_S(t)}{\langle S(t)\rangle} \right)\approx\frac{\delta_X(t)}{\langle S(t) \rangle}=\frac{X(t)}{\langle S(t) \rangle}
\end{eqnarray}
where we have used,
\begin{itemize}
    \item[(i)] $\vert\delta_S(t)\vert\ll\langle S(t) \rangle$, i.e.,  the variations  in $S(t)$ are much smaller than its mean value;
    
    \item[(ii)] $X(t)$ has, ideally, zero mean value;
    
    \item[(iii)] the term $\delta_X(t)\delta_S(t)/\langle S(t) \rangle^2$ can be neglected to a first order approximation.
\end{itemize}

Note the importance of good centralization between the beam and the PS detector when applying this approximation: a non-zero mean value of $X(t)$ would create a non-negligible term $\langle X(t)\rangle\delta_S(t)/\langle S(t)\rangle^2$. A direct consequence of (\ref{eq:approximation1}) is that the power spectra obtained from the division of $X(t)$ by the the mean value of $S(t)$ are approximately the same as obtained by the division of $ X(t)$ by the instantaneous values of $S(t)$. We refer to the former PSD as $\rm{PSD_{mean}}$ and to the latter as $\rm{PSD_{inst}}$.

Since dividing by a constant - namely, the mean value of $S(t)$ - does not affect the corner frequency of a Lorentzian function, which is proportional to the spring constant, one need only measure $X(t)$ and $Y(t)$ in order to obtain the radial trap stiffness. This is useful when using Data Acquisition boards (DAq) that have sampling frequency or data transfer rate limited by the number of active channels in the equipment. Also, it greatly reduces the computational cost of data analysis. 

Consider the following explicit models for $X(t)$ and $S(t)$,
\begin{eqnarray}
\label{eq:explicit_expressions}
    X(t) &=& \alpha_Xx_p(t)P(t)+\eta_{X}(t)\nonumber\\
    S(t) &=& \alpha_SP(t)+\eta_{S}(t)
\end{eqnarray}
where $x_p(t)$ is the particle position, $P(t)$ is the forward scattered power, $\eta_X(t)$ and $\eta_S(t)$ are the noise in the $X$ and $S$ channels and $\alpha_X, \alpha_S$ are proportionality constants. 
In this model, the oscillations in channels $X$ and $S$ due to particle motion are contained in the first term of each expression, while $\eta_X$ and $\eta_S$ are due to electronic noise in each channel, being independent of both the trapped particle and trapping laser. 
Plugging (\ref{eq:explicit_expressions}) into (\ref{eq:approximation1}) yields,
\begin{eqnarray}
\label{eq:position_x}
\frac{X(t)}{S(t)} &\approx&\frac{X(t)}{\langle S(t) \rangle} = \frac{\alpha_Xx_p(t)P(t)+\eta_X(t)}{\langle \alpha_S P(t)+\eta_S(t) \rangle}\approx \frac{\alpha_Xx_p(t)P(t)+\eta_X(t)}{\alpha_S\langle P(t)\rangle} \nonumber\\&\approx& \frac{\alpha_X\delta_{x_p}(t)[\langle P(t)\rangle +\delta_P(t)]+\eta_X(t)}{\alpha_S\langle P(t)\rangle}
\approx \frac{\alpha_X}{\alpha_S}x_p(t)+\frac{\eta_X(t)}{\alpha_S\langle P(t)\rangle}
\end{eqnarray}
where we have used,
\begin{itemize}
    \item[(i)] $\langle \eta_S(t)\rangle\ll\langle S(t) \rangle$, so that $\langle S(t)\rangle \approx \langle \alpha_S P(t) \rangle$;
    
    \item[(ii)] the mean value of the particle position $x_p(t)$ is zero in a harmonic trap;
    
    \item[(iii)] the product $\delta_x(t)\delta_P(t)$ can be neglected to a first order approximation, that is, $\vert\delta_P(t)\vert\ll\langle P(t) \rangle$.
\end{itemize}

Since the noise in the $X$ channel and the particle position are uncorrelated, the $\rm{PSD}$ can be separated in two parts. The first is an Aliased Lorentzian (AL), characteristic of discrete sampling of the particle position \cite{BergSorensen2004}. The second is a term proportional to the noise PSD in the $X$ channel. Moreover, this last terms is inversely proportional to the squared value of the trapping power, since $\alpha_S \langle P(t)\rangle$ is proportional to the latter. 
This provides a method for dealing with noisy detection systems: instead of fitting the $\rm{PSD}$ to a pure AL, one can measure the power spectrum of the noise in the $X$ channel when the trapping laser is off, which we call $\rm{PSD_{dark}}$, and fit the measured PSD to an AL added to a term proportional to $\rm{PSD_{dark}}$. This can also be easily extended to the PSD of motion in the axial direction, with the noise in the $X$ channel replaced by the noise in the $S$ channel. As a final remark, note that this is different from the whitening procedure used by \cite{Abbott2016} to deal with detectors that have frequency dependent response. 

One interesting alternative to a PS sensor is what we will call a \textit{knife-edge detector}. In such a setup, the collected scattered light is divided into two beams. One of the beams is focused on a regular photo-detector yielding a signal $S(t)$, proportional to the collected power $P(t)$. The other beam is partially blocked by a knife in such a way that when the particle is at the center of the trap, half of the light passes by the knife and is focused on a second photo-detector.

The movement of the particle in the $xy$-plane causes the beam to deviated by an angle proportional to the particle's position. The second detector thus gives a signal $X_k(t)$ containing two terms: one proportional to $P(t)$ relative to the detected power when the particle is not radially displaced; and another proportional to both $P(t)$ and $x_p(t)$.
In practice, due to electronic noise, the ratio between $X_k(t)$ and $S(t)$ reads,
\begin{eqnarray}
\label{eq:knife_x}
\frac{X_k(t)}{S(t)}&=&\frac{[\alpha_0+\alpha_X x_p(t)]P(t)+\eta_X(t)}{\alpha_SP(t)+\eta_S(t)}\approx\nonumber\\&\approx&\{[\alpha_0+\alpha_X x_p(t)]P(t)+\eta_X(t)\}\frac{1}{\alpha_SP(t)}\left(1-\frac{\eta_S(t)}{\alpha_S P(t)} \right)\approx\nonumber\\&\approx&\frac{\alpha_0}{\alpha_S}+\frac{\alpha_X}{\alpha_S}x_p(t)+\frac{1}{\alpha_SP(t)}\eta_X(t)-\frac{\alpha_0}{\alpha_S^2P(t)}\eta_S(t)\nonumber\\
&\approx&\frac{\alpha_0}{\alpha_S}+\frac{\alpha_X}{\alpha_S}x_p(t)+\frac{1}{\alpha_S\langle P(t)\rangle}\eta_X(t)-\frac{\alpha_0}{\alpha_S^2\langle P(t)\rangle}\eta_S(t)
\end{eqnarray}
In the above expression we have made the following approximations: 
\begin{itemize}
    \item[(i)] The amplitude of the electronic noise in the $S$ channel is much smaller than the total signal, that is, $\vert\eta_S(t)\vert\ll S(t)\approx \alpha_S P(t)$;
    
    \item[(ii)]The variations in $X_k(t)$ due to radial displacement and electronic noise are much smaller than $X_k(t)$, which means $\vert \alpha_Xx_p(t)P(t)+\eta_X(t)\vert \ll X_k(t)$ and implies that the product of these variations and $\eta_S(t)/\alpha_SP(t)$ can be neglected;
    \item[(iii)] The variations in $P(t)$ due to axial displacement are much smaller than $P(t)$, so that an expansion of the denominators for small $\delta_P(t)$ can be performed and second order terms involving products of $\delta_P(t)$ and $\eta_X(t) ,  \eta_S(t)$ can be neglected.
\end{itemize}

Since $x_P(t)$, $\eta_X(t)$ and $\eta_S(t)$ are all independent of each other, the PSD of $X_k(t)/S(t)$ - which we'll refer to as $\rm{PSD_{knife}}$ - is the sum of the power spectra of each of these signals weighted by different constants, with the constant term $\alpha_0/\alpha_S$ in (\ref{eq:knife_x}) making no contributions to the spectrum. 
This knife-edge setup allows one to measure the position of a trapped particle - and beam deviations in general - using detectors sensitive only to the light power. This is useful since not only PS detectors are often not readily available in a laboratory, but they also have limited bandwidth \cite{Huisstede2006}; with the knife-edge detector regular silicon detectors can be used in combination with low noise electronics when high frequencies must be accessed. This is an extension of the use of knife-edge detection, proposed by \cite{Braunsmann2014} in the field of high-speed atomic force microscopy, for situations in which the detected power varies with time.

As a final remark consider the expansion of $X_k(t)$:
\begin{eqnarray}
\label{eq:final}
X_k(t) &=& [\alpha_0+\alpha_X x_p(t)][\langle P(t) \rangle +\delta_P(t)]+\eta_X(t)\approx \nonumber\\ &\approx& \alpha_0\langle P(t) \rangle+\alpha_X\langle P(t) \rangle x_p(t)+\alpha_0 \delta_P(t) + \eta_X(t)
\end{eqnarray}
 Since movement in the radial and axial directions and the noise in the $X$ channel are independent, the PSD of $X_k(t)$ is given by the sum of the power spectra of the displacement in the $x$ and $z$ directions and the PSD of the noise in the $X$ channel. Hence fitting a sum of two AL and the spectrum of the noise to the measured PSD would in principle allow the measurement of trapping stiffness in the axial and radial directions by monitoring only one signal. The experimental feasibility of this calibration would rely on a large $\alpha_X/\alpha_0$ ratio.

The approximations used in the derivation of equations (\ref{eq:approximation1}), (\ref{eq:position_x}) and (\ref{eq:knife_x}) were not arrived at on theoretical grounds, but based on empirical evidence. 
Next we describe an experimental setup which implements the data analysis and knife-edge detection discussed in this section and justifies these approximations. 

\section{Experiment}
\begin{figure}[h!]
\centering
\includegraphics[scale = 1.2]{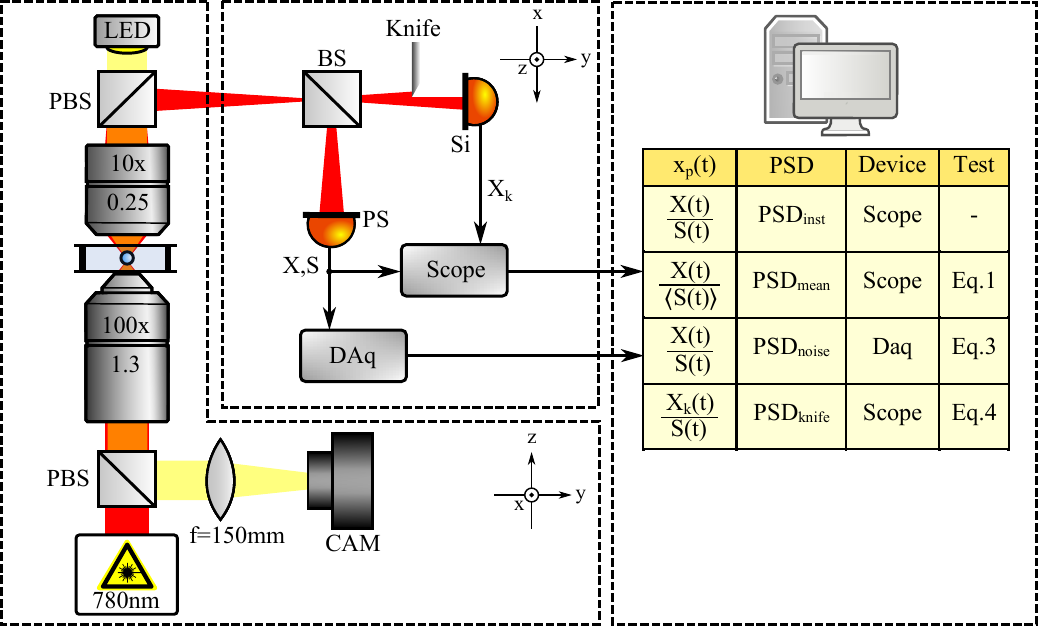}
\caption{Experimental setup used to test the derived approximations: a PS detector and a knife-edge detector are used to measure the beam deviation, while data is recorded by an oscilloscope and a DAq and used to calculate the PSD in four different ways.}
\label{fig:SetpTable}
\end{figure}

To test the validity of the above considerations for trapping stiffness measurements, an optical tweezer with multiple data acquisition channels was assembled, as in figure \ref{fig:SetpTable}.
A \SI{55}{mW} laser beam at \SI{780}{nm} (Toptica DL-pro) is focused by a high numerical aperture objective (Olympus UPlanFLN 100x, NA = 1.3) creating an optical trap for a \SI{1.15}{\mu m} silica bead (microParticles GmbH) in a water immersion. The scattered light from the particle is collected by a second objective lens (Olympus PlanN 10x, NA = 0.25) and divided into two beams by a beam splitter. 
The reflected light is collected by a PS detector (New Focus 2931) generating the signals $X(t)$ and $S(t)$, which are simultaneously monitored by an oscilloscope (sampling frequency $f_s=10$kHz) and a simple data acquisition board (DataQ 1100, sampling frequency $f_s=20$kHz). At the same time the transmitted beam is partially blocked by a knife and collected by a regular silicon photo-detector (Thorlabs DET100A2), creating the signal $X_k(t)$, also acquired by the oscilloscope.

\begin{figure}[h!]
\centering
\includegraphics[scale = 0.8]{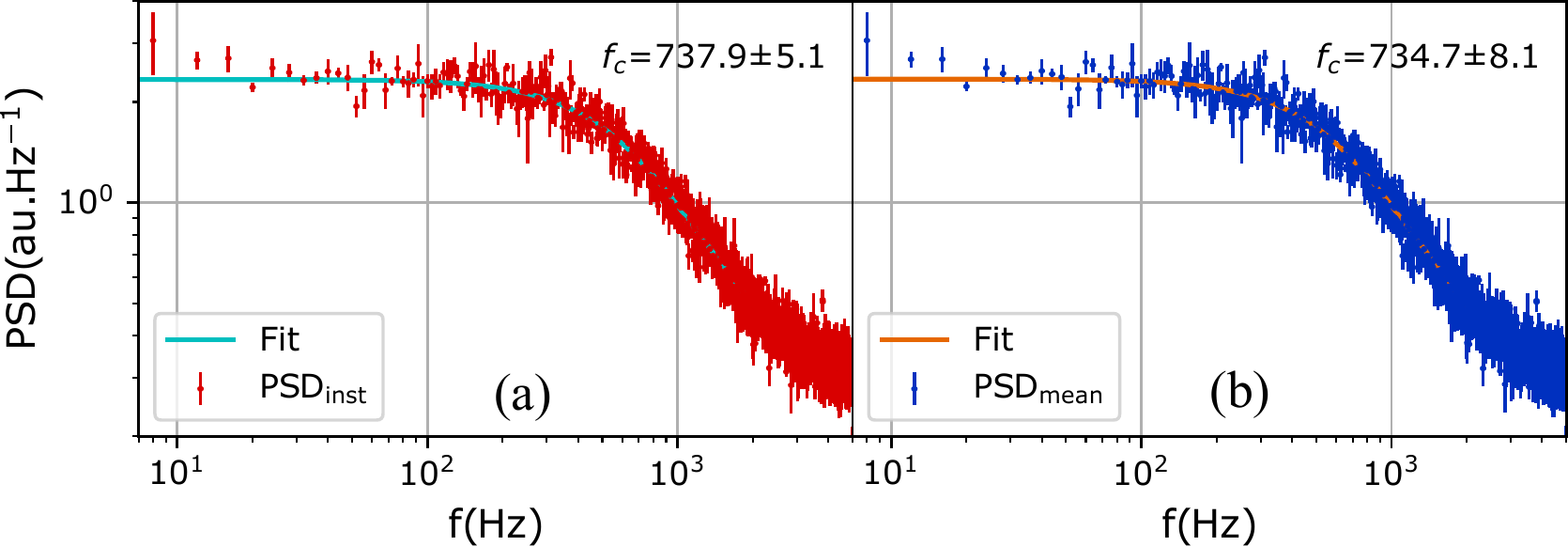}
\caption{Measured PSD and the resultant fit for (a) $\rm{PSD_{inst}}$ and (b) $\rm{PSD_{mean}}$.}
\label{fig:PSDmean}
\end{figure}

The data simultaneously recorded by the DAq and the oscilloscope was used to obtain the radial displacement of the trapped particle in four different ways, as summarized in the table from figure \ref{fig:SetpTable}. Due to the frequency independent electronic noise of the oscilloscope, an AL added to a constant value was used to fit $\rm{PSD_{inst}}$, resulting in figure \ref{fig:PSDmean}(a). Since this fitting procedure is described in the literature by \cite{Horst2010} in the context of position detection by video imaging, the corner frequency $f_c=737.9 \pm 5.1 \SI{}{Hz}$ of the fitted AL was taken to be the comparison standard for the other three methods. We do not attempt to translate the corner frequency into a spring constant value, since this would introduce errors relative to the medium's viscosity and temperature and to the particle's radius that would obscure the results of the intended comparison.

To test equation (\ref{eq:approximation1}), an AL added to a constant value was also fitted to the data from $\rm{PSD_{mean}}$, resulting in figure \ref{fig:PSDmean}(b). In this case a corner frequency of $f_c = 734.7\pm8.1 \SI{}{Hz}$ was obtained, in agreement to the standard value. 
This is consistent with the expectation that $\rm{PSD_{inst}}$ and  $\rm{PSD_{mean}}$ are almost identical and that the radial trap stiffness can be measured using either of them. 
Moreover, since dividing by a constant value does not change the corner frequency, one can obtain it directly from the spectral density of the signal $X(t)$ alone, reducing the amount of data processing required to measure trap stiffness in the radial direction. This enables real time measurements of the spring constant; one no longer needs to do the post-processing step of dividing $X(t)$ by $S(t)$.

\begin{figure}[h!]
    \centering
    \includegraphics[scale = 0.8]{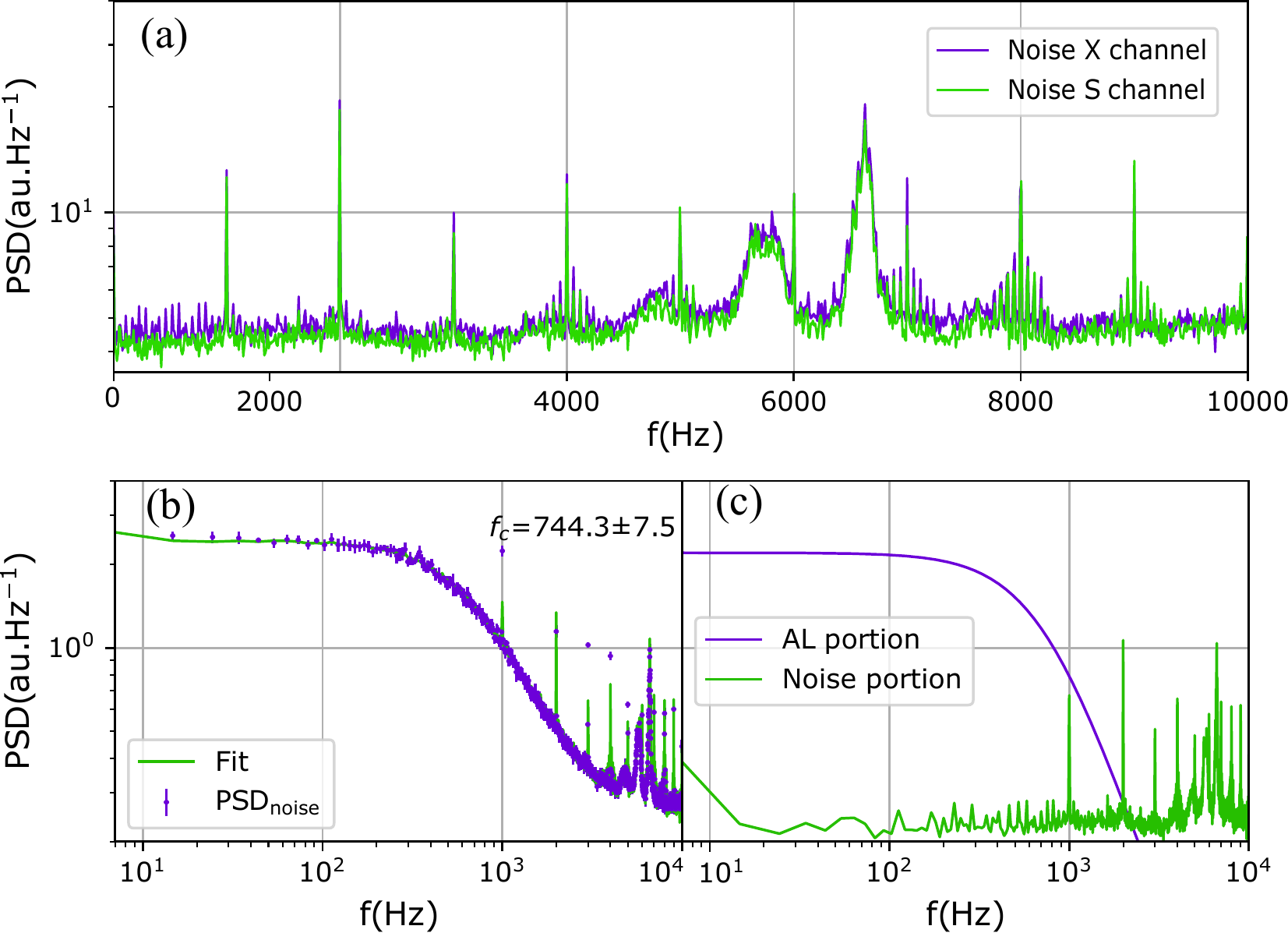}
    \caption{(a) PSD estimate for the electronic noise in the $X$ and $S$ channels. (b) Measured PSD and resultant fit for $\rm{PSD_{noise}}$. (c) AL and noise components of the resultant fit.}
    \label{fig:noise}
\end{figure}

The spectrum $\rm{PSD_{noise}}$ was used to test equation (\ref{eq:position_x}). The power spectrum $\rm{PSD_{dark}}$ of the electronic noise in the DAq's $X$ channel, measured when the laser was off, is displayed in figure \ref{fig:noise}(a). As it can be seen, it is not frequency independent, making an AL added to a constant value insufficient as the fitting function. Instead, as prescribed by equation \ref{eq:position_x}, we used an AL added to a term proportional to $\rm{PSD_{dark}}$. The result is shown in figure \ref{fig:noise}(b). A corner frequency of $f_c=744.3\pm7.5$ was found, which is also in agreement with the standard value. Figure \ref{fig:noise}(c) shows the Aliased Lorentzian and the noise portions of the fitted curve separately. 

The value of $\chi^2$ - with the residues weighted by the error in each point - was 4.4 times larger when using an AL added to a constant value instead of an AL added to $\rm{PSD_{dark}}$ to fit $\rm{PSD_{noise}}$. This corroborates with the result in equation $\ref{eq:position_x}$, that can be seen as a generalization of the prescription presented in \cite{Horst2010}, which deals with the particular case of white noise in the detection channel. 
With this result, one can use ``simple'' acquisition boards with complex electronic noise structure and yet perform quality measurements of the spring constant of an optical tweezer.

\begin{figure}[h!]
    \centering 
    \includegraphics[scale = 0.8]{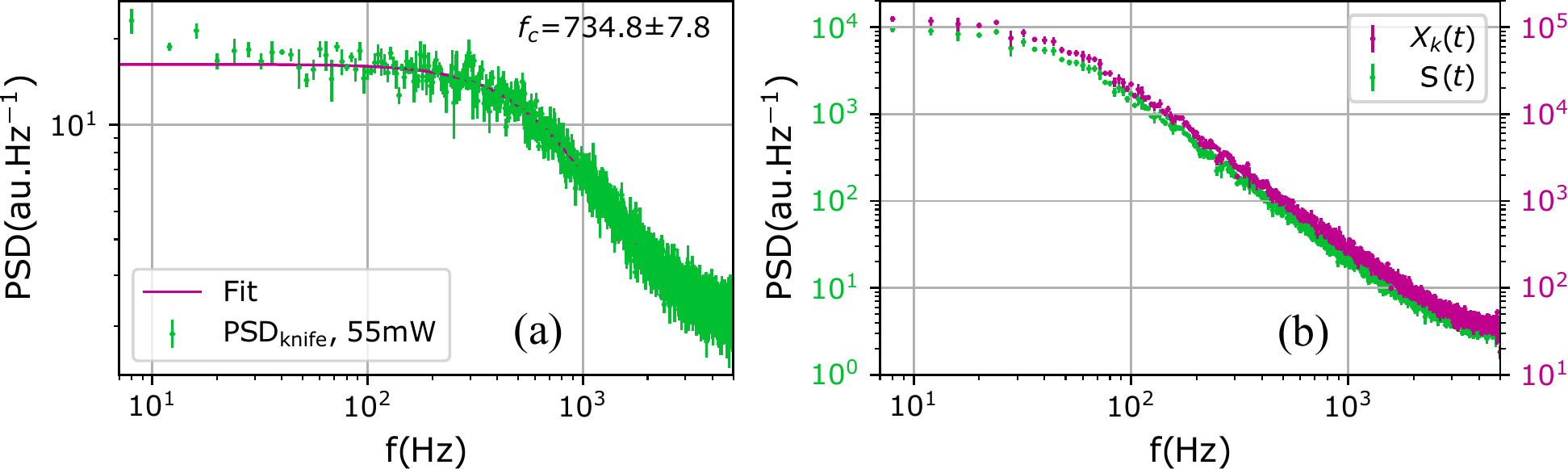}
    \caption{Knife-edge detection: (a) Measured $\rm{PSD_{knife}}$ and resultant fit; (b) PSD of $X_k(t)$ and $S(t)$ displayed separately.}
    \label{fig:PSD_knife}
\end{figure}

Finally, the knife-edge method was tested using $\rm{PSD_{knife}}$. Since the noise in each of the channels of the PS detector combined with the oscilloscope is taken to be an independent white noise, the third and fourth terms in (\ref{eq:knife_x}) results in a constant term in the PSD of the displacement in the $x$ direction. Therefore, $\rm{PSD_{knife}}$ was fitted to an AL added to a constant term, resulting in figure \ref{fig:PSD_knife}, for which $f_c=734.8\pm7.8$Hz. To avoid the low frequency noise introduced by the mechanical vibrations of the knife, the data bellow 30Hz was not included in the fit. This demonstrates that a regular silicon detector combined with a knife can indeed be used to measure a trapped particle's displacement in the radial direction. Moreover, axial displacements can be measured using the total value of the forwardly scattered power, while displacements in a radial direction orthogonal to the first one can be measured by dividing the beam once again, before it hits the total power detector, and partially blocking it with a knife positioned in a direction orthogonal to that of the first knife. This allows for full trap stiffness calibration using 3 regular silicon detectors. 

The necessity of dividing $X_k(t)$ by $S(t)$ when using the knife-edge method becomes clear when comparing figures \ref{fig:PSD_knife}(a) and \ref{fig:PSD_knife}(b), in which the PSD of $X_k(t)$ and $S(t)$ are displayed separately. The PSD of $X_k(t)$ clearly doesn't resemble $\rm{PSD_{knife}}$, being dominated by the PSD of $S(t)$, as suggested by equation (\ref{eq:final}) for the case in which the ratio $\alpha_X/\alpha_0$ is small.
 
\begin{figure}[h!]
\centering
\includegraphics[scale  = 0.8]{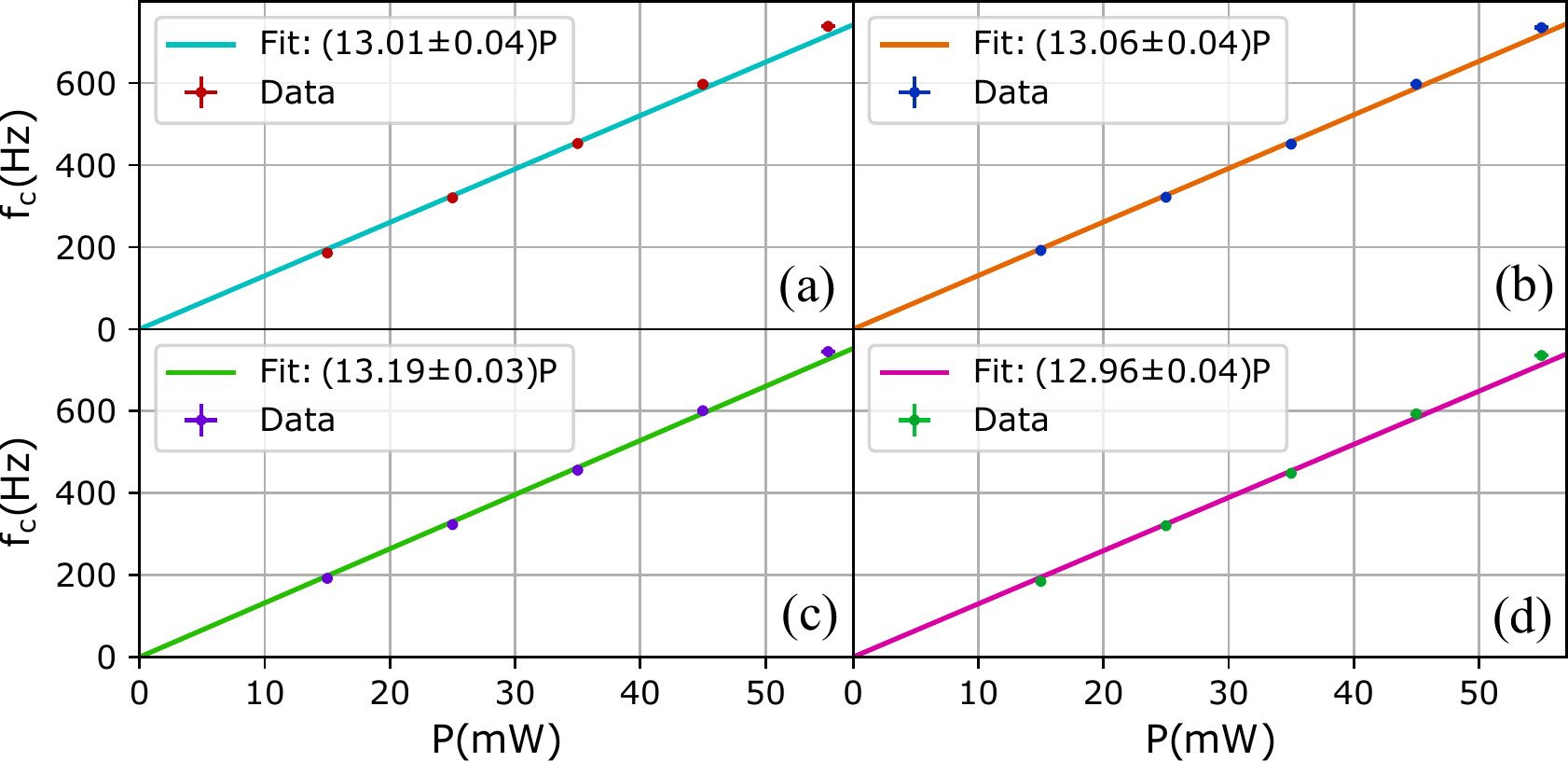}
\caption{Corner frequency as a function of trapping power extracted from (a) $\rm{PSD_{inst}}$; (b) $\rm{PSD_{mean}}$; (c) $\rm{PSD_{noise}}$ and (d) $\rm{PSD_{knife}}$ }
\label{fig:retas}
\end{figure}

Since the approximations made in the derivations of (\ref{eq:approximation1}), (\ref{eq:position_x}) and (\ref{eq:knife_x}) depend on the mean value of $S(t)$ being much larger than the variations in $S(t)$, it is expected some of them to be invalid as the trapping power gets small and $\eta_S(t)$ stops being negligible. To show the wide range of validity of the derived expressions, the four methods of corner frequency measurement were applied for trapping powers of 15mW, 25mW, 35mW and 45mW other than 55mW. The results for each method are displayed in figure \ref{fig:retas}. Since it is expected a linear dependence of $f_c$ on the trapping power - i.e $f_c = \alpha P$ - a straight line was used to fit the corner frequencies \cite{Jones2015}. The linear coefficients of each straight line, which are displayed in figure \ref{fig:retas}, shows that the expressions derived in this work are valid throughout the range of tested trapping powers.

\section{Conclusions}
Using suitable approximations, we have derived mathematical expressions for the position of a trapped particle measured by forward scattering detection. The first of these approximations imply that the PSD calculated from the division of $X(t)$ by $S(t)$ at each sample is almost identical to that obtained directly from $X(t)$. This is useful when dealing with acquisition devices that have a sampling frequency and data transfer rate limited by the number of active channels. Also, it reduces the need of post-processing, allowing for real time PSD visualization. 

The second derived expression resulted in a generalization of the procedure used in \cite{Horst2010}, in which a constant value is added to the Lorentzian function used to fit the experimental PSD. We concluded that for acquisition systems having frequency dependent electronic noise, the constant value can be substituted by the power spectral estimate of the noise in the $X$ channel, in the case of radial calibration, or in the $S$ channel, in the case of axial calibration. This, together with the whitening methods used in \cite{Abbott2016}, allow for the use of devices having complex response and noise spectral structure. 

Finally, the third expression suggested a method for using knife-edge detection in optical tweezers. Dividing the power detected after the knife by the total detected power, one can extend the use of this kind of detection to situations in which the beam power varies in time, in contrast to what was proposed in \cite{Braunsmann2014}. Since knife-edge detector can be implemented using detectors sensitive only to light power, such as regular silicon detectors, this can be useful when high frequencies are of interest. Also, we believe this kind of detector can be useful to the general optics community in situations in which position sensitive detectors are unavailable or when reducing costs and complexity is necessary \cite{Smith1999, Candia2013}.

All of the above expressions were experimentally tested and compared using a custom optical tweezer setup, equipped both with a position sensitive detector and the proposed knife-edge detector connected to two different acquisition devices. Good agreement between the three methods and the standard method was found when calculating the corner frequency of the PSD in the radial direction for different trapping powers. 

\section*{Acknowledgements}
The authors would like to thank Paulo Américo, Natan Viana and Luis Pires from UFRJ for useful discussions and advices during the development of this work. This work was supported by the
Serrapilheira Institute (grant number Serra-1709-21072), by Coordenac\~ao de
Aperfei\c{c}oamento de Pessoal de N\'ivel Superior - Brasil
(CAPES) - Finance Code 001 and by Conselho Nacional de Desenvolvimento Cient\'ifico e Tecnol\'ogico (CNPq).

\bibliography{main}

\begin{thebibliography}{10}

\bibitem{Ashkin1970}
A.~Ashkin.
\newblock Acceleration and trapping of particles by radiation pressure.
\newblock {\em Physical Review Letters}, 24(4):156--159, jan 1970.

\bibitem{Ashkin1986}
A.~Ashkin, J.~M. Dziedzic, J.~E. Bjorkholm, and Steven Chu.
\newblock Observation of a single-beam gradient force optical trap for
  dielectric particles.
\newblock {\em Optics Letters}, 11(5):288, may 1986.

\bibitem{Polimeno2018}
Paolo Polimeno, Alessandro Magazz{\`{u}}, Maria~Antonia Iat{\`{\i}}, Francesco
  Patti, Rosalba Saija, Cristian Degli~Esposti Boschi, Maria~Grazia Donato,
  Pietro~G. Gucciardi, Philip~H. Jones, Giovanni Volpe, and Onofrio~M.
  Marag{\`{o}}.
\newblock Optical tweezers and their applications.
\newblock {\em Journal of Quantitative Spectroscopy and Radiative Transfer},
  218:131--150, oct 2018.

\bibitem{Nussenzveig2017}
H.~Moys{\'{e}}s Nussenzveig.
\newblock Cell membrane biophysics with optical tweezers.
\newblock {\em European Biophysics Journal}, 47(5):499--514, nov 2017.

\bibitem{Dao2003}
M.~Dao, C.T. Lim, and S.~Suresh.
\newblock Mechanics of the human red blood cell deformed by optical tweezers.
\newblock {\em Journal of the Mechanics and Physics of Solids},
  51(11-12):2259--2280, nov 2003.

\bibitem{Zhang2019}
Zheng Zhang, Tom E.~P. Kimkes, and Matthias Heinemann.
\newblock Manipulating rod-shaped bacteria with optical tweezers.
\newblock {\em Scientific Reports}, 9(1), dec 2019.

\bibitem{Zhong2013}
Min-Cheng Zhong, Xun-Bin Wei, Jin-Hua Zhou, Zi-Qiang Wang, and Yin-Mei Li.
\newblock Trapping red blood cells in living animals using optical tweezers.
\newblock {\em Nature Communications}, 4(1), apr 2013.

\bibitem{Liu2018}
L.~R. Liu, J.~D. Hood, Y.~Yu, J.~T. Zhang, N.~R. Hutzler, T.~Rosenband, and
  K.-K. Ni.
\newblock Building one molecule from a reservoir of two atoms.
\newblock {\em Science}, 360(6391):900--903, apr 2018.

\bibitem{Hu2019}
M.-G. Hu, Y.~Liu, D.~D. Grimes, Y.-W. Lin, A.~H. Gheorghe, R.~Vexiau,
  N.~Bouloufa-Maafa, O.~Dulieu, T.~Rosenband, and K.-K. Ni.
\newblock Direct observation of bimolecular reactions of ultracold {KRb}
  molecules.
\newblock {\em Science}, 366(6469):1111--1115, nov 2019.

\bibitem{Monteiro2017}
Fernando Monteiro, Sumita Ghosh, Adam~Getzels Fine, and David~C. Moore.
\newblock Optical levitation of 10-ng spheres with nano- g acceleration
  sensitivity.
\newblock {\em Physical Review A}, 96(6), dec 2017.

\bibitem{Hebestreit2018}
Erik Hebestreit, Martin Frimmer, Ren{\'{e}} Reimann, and Lukas Novotny.
\newblock Sensing static forces with free-falling nanoparticles.
\newblock {\em Physical Review Letters}, 121(6), aug 2018.

\bibitem{Delic2020}
Uro{\v{s}} Deli{\'{c}}, Manuel Reisenbauer, Kahan Dare, David Grass, Vladan
  Vuleti{\'{c}}, Nikolai Kiesel, and Markus Aspelmeyer.
\newblock Cooling of a levitated nanoparticle to the motional quantum ground
  state.
\newblock {\em Science}, 367(6480):892--895, jan 2020.

\bibitem{Tebbenjohanns2020}
Felix Tebbenjohanns, Martin Frimmer, Vijay Jain, Dominik Windey, and Lukas
  Novotny.
\newblock Motional sideband asymmetry of a nanoparticle optically levitated in
  free space.
\newblock {\em Physical Review Letters}, 124(1), jan 2020.

\bibitem{Liu2012}
Zhirong Liu and Daomu Zhao.
\newblock Radiation forces acting on a rayleigh dielectric sphere produced by
  highly focused elegant hermite-cosine-gaussian beams.
\newblock {\em Optics Express}, 20(3):2895, jan 2012.

\bibitem{Jiang2011}
Yunfeng Jiang, Kaikai Huang, and Xuanhui Lu.
\newblock Radiation force of highly focused lorentz-gauss beams on a rayleigh
  particle.
\newblock {\em Optics Express}, 19(10):9708, may 2011.

\bibitem{Zhan2003}
Qiwen Zhan.
\newblock Radiation forces on a dielectric sphere produced by highly focused
  cylindrical vector beams.
\newblock {\em Journal of Optics A: Pure and Applied Optics}, 5(3):229--232,
  mar 2003.

\bibitem{Barry2018}
D.~A. Barry and J.-Y. Parlange.
\newblock Universal expression for the drag on a fluid sphere.
\newblock {\em {PLOS} {ONE}}, 13(4):e0194907, apr 2018.

\bibitem{XinCheng2001}
Yao Xin-Cheng, Li~Zhao-Lin, Guo Hong-Lian, Cheng Bing-Ying, and Zhang
  Dao-Zhong.
\newblock Effects of spherical aberration on optical trapping forces for
  rayleigh particles.
\newblock {\em Chinese Physics Letters}, 18(3):432--434, feb 2001.

\bibitem{Theofanidou2004}
Eirini Theofanidou, Laurence Wilson, William~J. Hossack, and Jochen Arlt.
\newblock Spherical aberration correction for optical tweezers.
\newblock {\em Optics Communications}, 236(1-3):145--150, jun 2004.

\bibitem{Dutra2012}
R.~S. Dutra, N.~B. Viana, P.~A.~Maia Neto, and H.~M. Nussenzveig.
\newblock Absolute calibration of optical tweezers including aberrations.
\newblock {\em Applied Physics Letters}, 100(13):131115, mar 2012.

\bibitem{Dutra2014}
R.~S. Dutra, N.~B. Viana, P.~A.~Maia Neto, and H.~M. Nussenzveig.
\newblock Absolute calibration of forces in optical tweezers.
\newblock {\em Physical Review A}, 90(1), jul 2014.

\bibitem{Alves2012}
P.~S. Alves and M.~S. Rocha.
\newblock Videomicroscopy calibration of optical tweezers by position
  autocorrelation function analysis.
\newblock {\em Applied Physics B}, 107(2):375--378, mar 2012.

\bibitem{BergSorensen2004}
Kirstine Berg-S{\o}rensen and Henrik Flyvbjerg.
\newblock Power spectrum analysis for optical tweezers.
\newblock {\em Review of Scientific Instruments}, 75(3):594--612, mar 2004.

\bibitem{Gibson2008}
Graham~M. Gibson, Jonathan Leach, Stephen Keen, Amanda~J. Wright, and Miles~J.
  Padgett.
\newblock Measuring the accuracy of particle position and force in optical
  tweezers using high-speed video microscopy.
\newblock {\em Optics Express}, 16(19):14561, sep 2008.

\bibitem{Drobczynski2012}
S{\l}awomir Drobczy{\'{n}}ski, Marcin Bacia, Marta Wo{\'{z}}niak, and Krzysztof
  Symonowicz.
\newblock Particle position measuring with optical tweezers using video
  processing.
\newblock In Jan Pe{\v{r}}ina, Libor Nozka, Miroslav Hrabovsk{\'{y}}, Dagmar
  Sender{\'{a}}kov{\'{a}}, Waclaw Urba{\'{n}}czyk, and Ondrej Haderka, editors,
  {\em 18th Czech-Polish-Slovak Optical Conference on Wave and Quantum Aspects
  of Contemporary Optics}. {SPIE}, dec 2012.

\bibitem{Huisstede2005}
J.~H.~G. Huisstede, K.~O. van~der Werf, M.~L. Bennink, and V.~Subramaniam.
\newblock Force detection in optical tweezers using backscattered light.
\newblock {\em Optics Express}, 13(4):1113, 2005.

\bibitem{Pralle1999}
A.~Pralle, M.~Prummer, E.-L. Florin, E.H.K. Stelzer, and J.K.H. Hörber.
\newblock Three-dimensional high-resolution particle tracking for optical
  tweezers by forward scattered light.
\newblock {\em Microscopy Research and Technique}, 44(5):378--386, mar 1999.

\bibitem{Jones2015}
Philip~H. Jones, Onofrio~M. Marago, and Giovanni Volpe.
\newblock {\em Optical Tweezers}.
\newblock Cambridge University Press, 2015.

\bibitem{Horst2010}
Astrid van~der Horst and Nancy~R. Forde.
\newblock Power spectral analysis for optical trap stiffness calibration from
  high-speed camera position detection with limited bandwidth.
\newblock {\em Optics Express}, 18(8):7670, mar 2010.

\bibitem{Smith1999}
Stephen~P. Smith, Sameer~R. Bhalotra, Anne~L. Brody, Benjamin~L. Brown,
  Edward~K. Boyda, and Mara Prentiss.
\newblock Inexpensive optical tweezers for undergraduate laboratories.
\newblock {\em American Journal of Physics}, 67(1):26--35, jan 1999.

\bibitem{Candia2013}
Carmen Noem{\'{\i}}~Hern{\'{a}}ndez Candia, Sara~Tafoya Mart{\'{\i}}nez, and
  Braulio Guti{\'{e}}rrez-Medina.
\newblock A minimal optical trapping and imaging microscopy system.
\newblock {\em {PLoS} {ONE}}, 8(2):e57383, feb 2013.

\bibitem{Grimm2012}
Matthias Grimm, Thomas Franosch, and Sylvia Jeney.
\newblock High-resolution detection of brownian motion for quantitative optical
  tweezers experiments.
\newblock {\em Physical Review E}, 86(2), aug 2012.

\bibitem{Abbott2016}
B.{\hspace{0.167em}}P.~Abbott et. al.
\newblock {GW}150914: First results from the search for binary black hole
  coalescence with advanced {LIGO}.
\newblock {\em Physical Review D}, 93(12), jun 2016.

\bibitem{Huisstede2006}
J.~H.~G. Huisstede, B.~D. van Rooijen, K.~O. van~der Werf, M.~L. Bennink, and
  V.~Subramaniam.
\newblock Dependence of silicon position-detector bandwidth on wavelength,
  power, and bias.
\newblock {\em Optics Letters}, 31(5):610, mar 2006.

\bibitem{Braunsmann2014}
Christoph Braunsmann, Veronika Prucker, and Tilman~E. Schäffer.
\newblock Optical knife-edge displacement sensor for high-speed atomic force
  microscopy.
\newblock {\em Applied Physics Letters}, 104(10):103101, mar 2014.

\end{thebibliography}

\end{document}